\documentclass[12pt]{article}
\usepackage[dvips]{graphicx}
\usepackage{pdproc}

  \makeatletter
  \def\@cite#1{[#1]}
  \makeatother
\begin{document}

\renewcommand{\thefootnote}{\alph{footnote}}

\title{ Interacting Dark Energy }

\author{ Xinmin Zhang}

\address{ Institute of High Energy Physics\\
Chinese Academy of Sceinces\\
P.O. Box 918(4)\\
Beijing 100049\\
P.R. China
\\ {\rm E-mail: xmzhang@mail.ihep.ac.cn}}

\abstract{ In this paper I discuss some of the phenomenologies of
models of the dark energy interacting with the ordinary matter.
After a very brief review about the current constraint on the
equation of the state of the dark energy from the SN and a new
scenario of dark energy {\it the Quintom}, I present models of
Quintessential Baryo(Lepto)genesis, Quintessino dark matter and
mass varying neutrinos in details. }

\normalsize\baselineskip=15pt

\section{Introduction}
In the concordance model of cosmology, $4 \%$ of the matter is
described by the standard model of the particle physics, however
the remaining $96 \%$ of matter ($23 \%$ the cold dark matter and
$73\%$ dark energy) are unknown. Understanding the dark sector of
the universe is a big challenge to the particle physics
(associated with the $4\%$ baryon matter is the question how
dynamically to generate the baryon number asymmetry).
 From the point of
view of particle physics, the leading candidates for cold dark matter are
the axion
and the neutralino.
The axion is a neutral spin-zero Pseudo-Goldstone boson associated with
the
spontaneous breaking of
the global $U_{PQ}(1)$ symmetry. Axions as
dark matter particles can be
produced by three different
mechanisms: vacuum alignment,
axion string decay and axion domain wall decay.

The neutralino is an electrically neutral hypothetical particle which
arises
in supersymmetric models. In many such models, e.g. in the MSSM (the
minimal
supersymmetric standard model),
the lightest supersymmetric particle (LSP) is stable, unless R-parity
violating interactions are
included. The LSP is generally thought to be the lightest neutralino
$\chi$.
The neutralinos in the Universe today are in general assumed to be a relic
of
an initially thermal neutralino distribution in the hot early Universe.
Based
on this thermal production mechanism, there have been many calculations of
the
LSP abundance as a function of
the MSSM parameters.
There, however exists a possibility of non-thermal production of
the neutralino. One of the example is the mechanism proposed by
Brandenberger, Jeannerot and me in
Ref.\cite{paper1} where the neutralinos are produced by the cosmic string
decays. Specifically we consider
models with an extra
$U(1)$ gauge symmetry in extensions of the MSSM. This $U(1)$
symmetry for instance could be
$U_{B-L}(1)$, where $B$ and $L$ are respectively baryon and lepton
numbers. Such models
explain the neutrino masses via the see-saw mechanism.

The basic idea of our mechanism is as follows. When the extra $U(1)$
symmetry
which we have introduced gets broken at a scale $\eta$, a network of
strings is produced by the usual Kibble mechanism.
 If, as we assume, the fields excited in the
strings couple to the neutralino $\chi$, then a non-thermal
distribution of $\chi$ particles will be generated during the
process of string decay. Because of the non-thermal distribution
the neutralinos for some cases behave like "warm" dark matter
which as I in collaboration with Lin, Huang and Brandenberger have
shown in Ref.\cite{paper2} provides a solution to the possible
problem of the cold dark matter at the subgalactic scale.
Furthermore  in the mechanism of non-thermal production the relic
abundance of the dark matter particles are not "directly"
correlated with the "interacting" rate of the dark matter
particles with the ordinary matter. This helps enhance the
detection rate in the search for dark matter
particles\cite{paper3}.

In the literature there have been many proposals for the non-thermal
production of neutralinos. In the models with split supersymmetry the
neutralino dark matter particles are shown recently to be
produced non-thermally\cite{paper4}.

Regarding dark energy, the recent data from type Ia supernovae
and cosmic
microwave background (CMB) radiation have provided strong
evidences for a spatially flat and
accelerated expanding universe at the present time.
In the context of
Friedmann-Robertson-Walker cosmology, this acceleration is attributed to
the domination of a component, dubbed dark energy.
The simplest candidate for dark energy
seems to be a remnant small cosmological constant. However,
many physicists are attracted by
the idea that dark energy is due to a dynamical component, such as a
canonical scalar field $Q$,
named {\it Quintessence}.

Being a dynamical component, the scalar field dark energy is
expected to interact with the ordinary matters. There are many
discussions on the explicit couplings of quintessence to baryons,
dark matter and photons, however as argued in Refs.
\cite{paper5,paper6} for most of the cases the couplings are
strongly constrained. But there are exceptions. For example,
Carroll \cite{paper5} has considered an interaction of form $Q
F_{\mu\nu}{\tilde F^{\mu\nu}}$ with $F_{\mu\nu}$ being the
electromagnetic field strength tensor which has interesting
implication on the rotation of the plane of polarization of light
coming from distant sources. Recent data on the possible variation
of the electromagnetic fine structure constant has triggered
interests in studies related to the interactions between
quintessence and the matter fields.

This paper is organized as follows: In section II, I will briefly
review on the current constraint on the dark energy and present a
new scenario of dark energy, the Quintom which I will show is
consistent with and is mildly favored by the recent SN Ia data; In
section III, I review Quintessential Baryo(Lepto)genesis; In
section IV, I consider a scenario of supersymmetric version of a
Quintessence model with interactions between the Quintessence and
the ordinary matter and present a model of unifying the dark
matter and dark energy; In section V, I consider the possibility
of neutrino interacting with the dark energy scalars and the idea
of mass varying neutrinos;
 Section VI is my conclusion of
the paper.

\section{Supernova Constraint on the
Equation of State of Dark Energy}

Despite the current theoretical ambiguity for the nature of dark
energy, the prosperous observational data ({\it e.g.} supernova,
CMB and large scale structure data and so on ) have opened a
robust window for testing the recent and even early behavior of
dark energy using some parameterizations for its equation of
state. The recent fits to the SN data and CMB etc in the literature find
that
the behavior of
dark energy is to great extent in consistency with a cosmological
constant. However, when the equation of state is not
restricted to be a constant, the fit to observational data
improves dramatically. For example, Huterer
and Cooray \cite{paper7} produced uncorrelated and nearly
model-independent band power estimates of the equation of state of dark
energy and its density as a function of redshift, by fitting to
the recent SNe Ia data they found marginal (2-$\sigma$) evidence
for $W(z) < -1$ at $z < 0.2$.

In Ref.\cite{paper8}
using the low limit of cosmic ages from globular cluster and the
white dwarfs: $t_0 > 12$Gyr, together with recent new high
redshift supernova observations from the HST/GOODS program and
previous supernova data, we give a considerable estimation of the
equation of state for dark energy, with uniform priors as weak as
$0.2<\Omega_m<0.4$ or $0.1<\Omega_m h^2<0.16$. We find cosmic age
limit plays a significant role in lowering the upper bound on the
variation amplitude of dark energy equation of state, and
phenomenologically an equation of state of the dark energy which transits
from below -1 to above -1 as the redshift increases is mildly favored.
If such a result holds on with the accumulation
of observational data, this would be a great challenge to current
cosmology.
 Firstly, the cosmological constant as a
candidate for dark energy will be excluded and dark energy must be
dynamical. Secondly, the simple dynamical dark energy models
considered vastly in the literature like the
quintessence or the
phantom can not be
satisfied either.

In the quintessence model, the energy density and the pressure for
the quintessence field are
\begin{equation}
 \rho=\frac{1}{2}\dot Q^2+V(Q)~,~~p=\frac{1}{2}\dot Q^2-V(Q)~.
\end{equation}
 So, its equation of state $W=p/\rho$ is in the range
$-1\leq
W\leq 1$ for $V(Q)>0$. However, for the phantom
which has the opposite sign of the kinetic term compared with the
quintessence in the Lagrangian,
\begin{equation}
 \mathcal{L}=-\frac{1}{2}\partial_{\mu}Q\partial^{\mu}Q-V(Q)~,
 \end{equation}
the equation of state $W=(-\frac{1}{2}\dot
Q^2-V)/(-\frac{1}{2}\dot Q^2+V)$ is located in the range of $W\leq
-1$. Neither the quintessence nor the phantom alone can fulfill
the transition from $W>-1$ to $W<-1$ and vice versa. In
Ref.\cite{paper8} we have proposed a new scenario of the dark
energy, dubbed {\it Quintom}. A simple Quintom model consists of
two scalar fields, one being the quintessence with the other being
the phantom field. This type of Quintom model will provide a
scenario where at early time the quintessence dominates with
$W>-1$ and lately the phantom dominates with $W$ less than $-1$,
satisfying current observations. A detailed study on the
cosmological evolution of this class of Quintom model is performed
in Ref.\cite{paper9}. The Quintom models are different from the
quintessence or phantom in the determination of the evolution and
fate of the universe. Generically speaking, the phantom model has
to be more fine tuned in the early epochs to serve as dark energy
today, since its energy density increases with expansion of the
universe. Meanwhile the Quintom model can also preserve the
tracking behavior of quintessence, where less fine tuning is
needed. In Ref.\cite{paper10} we have studies a class of Quintom
models with an oscillating equation of state and found that
oscillating Quintom can unify the early inflation and current
acceleration of the universe, leading to oscillations of the
Hubble constant and a recurring universe. Our oscillating Quintom
would not lead to a big crunch nor big rip. The scale factor keeps
increasing from one period to another and leads naturally to a
highly flat universe. The universe in this model recurs itself and
we are only staying among one of the epochs, in which sense the
coincidence problem is reconciled.

In addition to the Quintom model mentioned above there are at
least two more possibilities in the Quintom model buildings.
 One will be the scalar field models with
non-minimal coupling to the gravity
where the effective equation of the state can be arranged to
change from above -1 to below -1 and vice versa. For a single
scalar field coupled with gravity minimally, one may consider a
model with a non-canonical kinetic term with the following
effective Lagrangian \cite{paper8}:
 \begin{equation}
\mathcal{L}=\frac{1}{2}f(T)\partial_{\mu}Q\partial^{\mu}Q-V(Q)~,
 \end{equation}
where $f(T)$ in the front of the kinetic term is a dimensionless
function of the temperature or some other scalar fields. During
the evolution of the universe when $f(T)$ changes sign from
positive to negative it gives rise to an realization of the
interchanges between the quintessence and the phantom scenarios.

\section{Quintessential Baryo(Lepto)genesis}

In this section I will review Quintessential
Baryo(Lepto)genesis \cite{paper11}.
Consider
 a
type of interaction of Quintessence with the matter,  which
 in terms of an effective lagrangian is given by
\begin{eqnarray}\label{lagr}
{\cal L}_{eff}=\frac{c}{M }{\partial_{\mu}Q } ~J^{\mu},
\end{eqnarray}
where $M$ is the cut-off scale which for example could be the Planck
mass $M_{pl}$ or the scale of Grand Unification Theory $M_{GUT}$,
and $c$ is the coupling constant which characterizes
the strength of Quintessence interacting with the ordinary matter in
the Standard Model of the electroweak theory.

Specifically we take in Ref.\cite{paper11} $J^{\mu}$ to be the baryon
current
$J^{\mu}_B$ or the current of baryon number minus lepton
number
$J_{B-L}^{\mu}$,
and study their implications on baryogenesis or leptogenesis. The
lagrangian in
Eq.(\ref{lagr})
involves derivative and
obeys the symmetry $Q \rightarrow Q + {\it constant}$, so
the Quintessence potential will not be modified by the quantum
corrections.

The mechanism of generating the baryon number asymmetry in this scheme
follows closely the
spontaneous baryogenesis\cite{paper12}.  The term in Eq.(\ref{lagr}), when
$\dot Q$
is non-zero during the evolution of spatial flat
Friedmann-Robertson-Walker Universe
violates CPT invariance and
generates an effective chemical potential $\mu_b$ for baryons, ${\it
i.e.}$,
\begin{eqnarray}
& &{c\over{M }}\partial_{\mu}Q
J^{\mu}_B \rightarrow ~c ~
{\dot{Q}\over{M }}n_{B}= ~c ~{\dot{Q}\over{M }}(n_{b}-n_{\bar
b})~,\nonumber\\
& &\mu_{b}=-~c ~~{\dot{Q}\over{M }}=-\mu_{\bar b}~.
\end{eqnarray}
In thermal equilibrium the baryon number asymmetry is given by (when
$T\gg m_{b}$)
\begin{eqnarray}
n_{B}=\frac{g_{b}T^{3}}{6}({\mu_{b}\over T}+{\cal O}({{\mu_{b}\over
T})}^3)\simeq c \frac{g_{b}\dot Q T^{2}}{6 M }~,
\end{eqnarray}
where $g_b$ counts the internal degree of freedom of the baryon.
Using the familiar expression for entropy density
\begin{equation}
s =\frac{2 \pi^2}{45} g_{\star} T^3,
\end{equation}
we arrive at the final expression for the baryon to entropy ratio
\begin{equation}\label{bnumber}
n_{B}/s\simeq \frac{15 c}{4 \pi^2}\frac{g_{b}\dot Q}{g_{\star}M  T}~.
\end{equation}
$\dot Q$ in Eq.(\ref{bnumber}) can be obtained by solving
 the equation of motion of
Quintessence given below
\begin{equation}\label{motion}
\ddot Q+3H\dot Q+V^{'}(Q)=-{ c\over M }(\dot n_{B}+3Hn_{B})~,
\end{equation}
where H is the Hubble constant and $V(Q)$ is the potential of
Quintessence field.
For the radiation dominated era the Hubble constant is
\begin{equation}
H={1\over 2t}=1.66 g^{1/2}_{\star}\frac{T^2}{M_{pl}}~.
\end{equation}
The right-handed side of Eq.(\ref{motion}) is about
$-\frac{c g_b} {6} \frac{T^2 }{M^2 }(\ddot Q+H\dot Q)$
and can be neglected unless in the
very early universe when the temperature $T$ is close to the cut-off
scale.

In Ref.\cite{paper11}we consider a model which has the tracking property,
where
the potential has a modified exponential form\cite{paper13},
\begin{equation}\label{potential}
V(Q)=f(Q)e^{-{\lambda\over m_{pl}}Q}~.
\end{equation}
Then we obtain
a ratio of the baryon number to entropy given by
\begin{eqnarray}\label{bnumber2}
{n_{B}\over s}|_{T_D}\sim 0.01  c \frac{T_D}{M }~,
\end{eqnarray}
where $T_D$ denotes the epoch when the B-violating interactions
freeze out. In the standard model of particle physics, the baryon
number is violated by the Sphaleron processes with $T_{D}$ around
100 GeV, so with $c \sim {\cal O}(1)$ to have $n_B / s \sim
10^{-10}$ it requires $M \leq 10^{10}{\rm GeV}$ which will be
possible in models for instance with large extra dimension. For a
large value of $M$, such as $M= M_{pl}$ or $M=M_{GUT}$, $T_D$
needs also to be large and in general can be achieved in GUT
easily. However, if the B-violating interactions conserve $B-L$,
the asymmetry generated will be erased by the electroweak
Sphaleron. So now we turn to leptogenesis \cite{paper14}. We take
$ J^{\mu}$ in Eq. (4) to be $J_{B-L}^\mu$. Doing the calculations
with the same procedure as above for $J^{\mu} = J^{\mu}_{B}$ we
have the final asymmetry of the baryon number minus lepton number
\begin{eqnarray}\label{fnumber2}
 {n_{B-L}\over s}|_{T_D}\sim 0.1  c \frac{T_D}{M }.
\end{eqnarray}
The asymmetry $n_{B-L}$ in (\ref{fnumber2})  will be converted to baryon
number
asymmetry when electroweak
Sphaleron $B+L$
interaction is in thermal equilibrium which happens for
temperature in the range of $10^2 ~{\rm GeV}
\sim 10^{12}{\rm GeV}$. $T_D$ in (\ref{fnumber2}) is the temperature below
which the $B-L$ interactions freeze out.

This type of mechanism for baryo(lepto)genesis has been generalized in
models of k-essence\cite{paper15} and others\cite{paper16}.

One silent feature of this scenario for baryo(lepto)genesis is
that the present accelerating expansion and the
generation of the matter and antimatter
asymmetry of our universe
is described in a {\it unified} way. In this scenario the baryon number
asymmetry is generated in
thermal equilibrium which violates one of the conditions by
Sakharov. This is due to the existence of the
CPT violating {\it Ether}
during the evolution of the quintessence scalar field.

In the traditional version of the leptogenesis the heavy
right-handed neutrinos are introduced and their non-thermal
equilibrium decays, coupled with the electroweak sphaleron
process, generate the required baryon number asymmetry. In general
at least two types of the right-handed neutrinos are needed for a
successful leptogenesis. In the quintessential leptogenesis, there
is no need to introduce the heavy right-handed neutrinos and a
degenerated-mass pattern for the light neutrino is
favored\cite{paper11,paper15}.

One may wonder if this type of CPT violation will affect the
laboratory experiments.
At present time the quintessence field is slowly rolling and $\dot
Q$ is bounded from above. To get the maximal value,
$\dot Q_c$, note that
$\frac{1}{2} {\dot
Q}^2 \leq \rho_Q \leq \rho_c \sim 10^{-47} ~[ {\rm GeV}]^4$. So we have
$\dot Q_c \leq 10^{-23} ~[ {\rm GeV} ]^2$.
The experiment of CPT test with a spin-polarized torsion
pendulum \cite{paper17} puts strong limits on
the axial vector background $b_\mu$ which
is defined by ${\cal L}=b_{\mu}{\bar e} \gamma^\mu \gamma_5 e$
\cite{paper18}:

\begin{equation}
 |{\vec b}| \leq 10^{-28} ~{\rm GeV}~.
\end{equation}
For the time component $b_0$, the bound is relaxed to be at the
level of $10^{-25}$ GeV \cite{paper19}. Taking the current $J^\mu$
in Eq. (\ref{lagr}) to be ${\bar e} \gamma^\mu \gamma_5 e$, $b_0$
here corresponds to $c \frac{\dot Q}{M}$ and it requires that
$b_0\sim c ~ 10^{-23} \frac{ {\rm GeV}^2}{M}\leq 10^{-25}~{\rm
GeV}$. This puts a constraint on the cutoff scale $M$, however, if
taking $M$ to be around the Planck or GUT scale the CPT violating
effects at the present time is much below the current experimental
sensitivity. In Ref.\cite{paper20} we have pointed out that this
type of phenomenon associated with CPT violation can be tested in
the future cosmic microwave background (CMB) polarization
experiments, such as PLANCK and CMBpol.

In the examples we considered above the dark energy is given by a
dynamical scalar field. If the dark energy is simply the vacuum
energy we can replace the $\partial_\mu Q /M$ in (4) by
$\partial_\mu f(R)$ where $f(R)$ a dimensionless function of the
Ricci scalar $R$. This kind of mechanism for baryogenesis dubbed
gravitational baryogenesis was first proposed in
Ref.\cite{paper21}. In Ref.\cite{paper21} $f(R)=R/M^2$ is taken,
however the Einstein equation, $R=8\pi G T^{\mu}_{\mu}=8\pi G
(1-3w)\rho$, tells us that $\dot f(R)=0$ in the radiation
dominated epoch of the standard Friedmann-Robertson-Walker (FRW)
cosmology. The authors of Ref. \cite{paper21} have considered
three different possibilities of obtaining a non-vanishing $\dot
R$ which include the effects of trace anomaly, reheating and
introducing a non-thermal component with $w>1/3$ dominant in the
early universe. In the braneworld scenario Shiromizu and Koyama in
Ref. \cite{paper22} provided another example for $\dot R\neq 0$.
In Ref.\cite{paper23} we take $f(R)\sim \ln R$, and propose a new
model of gravitational leptogenesis. We explicitly show in
\cite{paper23} that the term $\partial_{\mu}f(R)\sim
\partial_{\mu}R/R$ does not vanish during the radiation dominated
epoch and the observed baryon number asymmetry can be generated
naturally via leptogenesis\cite{paper23}.

\section{ Quintessino as Dark Matter Particle}

There have been many proposals for the candidates of dark matter or dark
energy, however it is always interesting to have a single theory
which explains both. In this section I will present a model where
the components of the dark matter (Quintessino) and the dark
energy (Quintessence) belong to one superfield, which is
very much like the quarks and leptons in the same group representation in
the
grand unified theories.
In this scenario,  the present acceleration of the Universe is driven by
the
dynamics of Quintessence
and, at the same time, the superpartner of Quintessence, the Quintessino,
makes up the dark matter of the Universe.

From the point of view of particle physics, fundamental interactions,
as widely believed, may be supersymmetric (SUSY)
beyond the TeV scale.
 In a SUSY theory, the
Quintessence boson will be accompanied by a 2-component neutral fermion
($\tilde{Q}$), {\it Quintessino}, and a scalar ($\sigma_q$), the
Squintesson.
Note that the current observational data indicate
that
the potential of the Quintessence field
around present epoch should be very flat and consequently its
effective mass will be extremely small, $m_Q\leq
H_0\sim 10^{-33}$ eV. The Quintessence behaves like a
pseudo-Goldstone boson, so one would expect
its fermionic superpartners also light. A naive
dimensional analysis in a model
independent way indicates $m _{\tilde{Q}}
\sim \mathcal{O}(M_{SUSY}^2/ \Lambda )$, where $\Lambda$ corresponds to
the decay constant of the Pseudo-Goldstone boson, the Quintessence here.
It may be possible, however, in the similar way as for
axino and Majorino\cite{paper24} that the Quintessino
receives a large mass in a specific model.

In the minimal supersymmetric standard model (MSSM) with a
conserved R-parity, the lightest SUSY particle (LSP), taken
usually as the lightest neutralino, $\chi$, is stable and serves
as an ideal candidate for dark matter. However, if the Quintessino
is lighter than $\chi$, the neutralino could decay and here we
study the possibility that the $\chi$ decay product, the
Quintessino, forms dark matter of the Universe.

In terms of an effective lagrangian we
in Ref.\cite{paper25} introduce new interactions
between Quintessence and ordinary matter. We impose
the shift symmetry, $Q \rightarrow Q + C$, which implies that the
interactions
of
the Quintessence with matter should involve derivatives.
In general there are two classes of
operators at dimension 5, one with fermions $f$ \cite{paper11} and the
other
one
with Higgs boson $H$ of the standard electroweak theory
\begin{eqnarray}
\label{qff}
\mathcal{L}_{Qff}&=&\frac{ 1 }{\Lambda}\partial_\mu Q
( c_{ij}^R \bar{f_i}_R\gamma^\mu f_{j R}
+ c_{ij}^L \bar{f_i}_L\gamma^\mu f_{j L}) \ ,\\
\label{qhh}
\mathcal{L}_{QHH}&=&\frac{c_H}{\Lambda} i\partial_\mu Q \left(
H^\dagger D^\mu H - (D^\mu H)^\dagger H
\right)\ ,
\end{eqnarray}
where $\Lambda$ represents the cutoff energy scale and $D^\mu$ is
the gauge covariant derivative. Several constraints are set on the
cutoff scale $\Lambda$. First, since the Quintessence is very
light, the coupling in the forms above will lead to an energy-loss
channel for stars. The cutoff is bounded below, $\Lambda > 2\times
10^9 GeV$, in order not to lead conflict with the observational
limits on the stellar-evolution time scale. The SN 1987A
observation also constrains this ``invisible channel'' and leads
to $\Lambda > 6\times 10^9 GeV$. The interactions in Eq.
(\ref{qff}) for $i = \mu$ and $j = e$ also induce lepton flavor
changing decay $ \mu \rightarrow e + Q$ with the branching ratio
given by Br$(\mu\to e Q)= \frac{3\pi^2}{\Lambda^2}\frac{1}{(m_\mu
G_F)^2}$ for $c_L^{e\mu}=c_R^{e\mu}=1$. The familon search
experiments set the bound on the cutoff scale as $\Lambda >
4\times 10^9 GeV$. The operator in Eq. (\ref{qhh}) gives rise to a
mixing between the Quintessence and the gauge boson $Z_{\mu}$,
which induces an effective coupling of the Quintessence to the
light fermions\cite{paper26}. The astrophysical experiments put a
limit $\Lambda > 3\times 10^9 GeV$. In a word, the present
astrophysical and laboratory experimental limit on the energy
scale $\Lambda$ of an axion-like pseudoscalar coupling
with matter is around $ 
10^{10} GeV$.

To calculate the decay rate of the neutralinos into quintessino,
we supersymmetrize the interactions above by introducing
the gauge and supersymmetric invariant Lagrangian
\begin{equation}
\label{susylag}
\mathcal{L}= \frac{c}{\Lambda}
\hat{Q} \Phi^\dagger e^{2gV} \Phi|_{\theta\theta\bar{\theta}\bar{\theta}}
+ h.c.\ ,
\end{equation}
where $\hat{Q}=(\sigma_q+ i Q) +\sqrt{2} \theta \tilde{Q} +\theta\theta F$
is
the chiral superfield containing Quintessence $Q$ and its fermionic
partner $\tilde{Q}$, $\Phi$ is any matter superfield in the
MSSM and $V$ is the vector superfield. We notice that this Lagrangian
possesses
the shift symmetry, i.e., $\hat{Q}\to \hat{Q} + i\Lambda C $.
When expressing it in terms of the component fields, we obtain
the needed couplings in Eqs. (\ref{qff}) and (\ref{qhh}).

Given the constraints on the effective operators above by
the current astrophysical and laboratory experimental data,
we have studied in Ref.\cite{paper25} the implications in Quintessino dark
matter.
Our results show that Quintessino can be a good candidate for CDM or
WDM through the late-time decay of the NLSP of ordinary superparticle,
which can be the Bino, the sleptons or Higgsino-like neutralino.

Generally the quintessino dark matter particles interact with the ordinary
matter very weakly, so the traditional techniques for the direct or
indirect detection of the neutralino dark matter
would not be useful for detecting quintessino dark matter.
There are, however, several silent features of the quintessino
dark matter model which make it distinguishable.
Firstly, we should note that being one superfield the quintessino
has the same interaction with matter as the quintessence and thus be
severely constrained.

Secondly, the BBN observation indicates that the light element
$^7$Li may be underabundant compared with the theoretical
estimate. Decaying particles after BBN might provide one way to
solve the problem. As shown explicitly in Ref. \cite{paper25} (and
argued for the general case of SuperWIMP such as gravitino dark
matter models in Ref. \cite{paper27}) the electromagnetic energy
associated with the non-thermal production of quintessino dark
matter particles can play the role to reconcile the observations
and the theory.

Thirdly, the quintessino dark matter in this model is produced
nonthermally. The property  of the quintessino dark matter is
characterized by the comoving free streaming scale.
Depending on the time when the NLSP decay and the initial
energy of the quintessino when it is produced,
it can be either cold or warm  dark matter.
In the latter case it helps solve the problems of the
cold dark matter on subgalactic scales.

Fourthly, this scenario for dark matter predicts the existence of
long-lived NLSP with life time $10^5-10^8$ sec. For the stau NLSP
it can be produced and collected on colliders and the properties
of quintessino can be studied by examining the stau decay. The
scenario with gravitino LSP and stau NLSP is studied in the
literature\cite{paper28}. For the case of quintessino LSP the stau
will have different decay modes. In Ref.\cite{paper29} we have
studied the possibility of detecting $\tilde{\tau}$ produced by
the high energy cosmic neutrinos interacting with the earth
matter. By a detailed calculation we find that the event rate is
one to several hundred per year at a detector with effective area
of $1 km^2$.

\section{Mass Varying Neutrinos}
There are at least two observations to motivate the speculations
on the connections between the neutrinos and the dark energy: 1)
in models of dark energy with a remnant small cosmological
constant or, true or false vacuum energy $\rho \sim {( 2 \times
10^{-3} {\rm ev} )}^4$. This energy scale $\sim 10^{-3}$ ev is
smaller than the energy scales in particle physics, but
interestingly is comparable to the neutrino masses; 2) in
Quintessence-like models $m_Q \sim 10^{-33}$ eV, which
surprisingly is also connected to the neutrino masses {\it via} a
see-saw formula $m_Q \sim {m_\nu^2 / m_{pl}}$ with $m_{pl}$ the
planck mass.

Is there really any connections between the neutrinos and dark
energy? Given the arguments above it is quite interesting to make
such a speculation on this connection. If yes, however in terms of
the language of the particle physics it requires the existence of
new dynamics and new interactions between the neutrino and the
dark energy sector. Recently there are some studies in the
literature on the possible realization of the models on neutrinos
and dark
energy\cite{paper30,paper11,paper15,paper31,paper32,paper33,paper34,paper35,paper36}.

Qualitatively these models have made at least two interesting
predictions: 1) neutrino masses are not constant, but vary during
the evolution of the universe; 2) CPT is violated in the neutrino
sector due to the CPT violating {\bf Ether} during the evolution
of the Quintessence scalar field\cite{paper11,paper15}.
Quantitatively these predictions will depend on the dynamics
governing the coupled system of neutrino and dark energy.

One of the possible couplings\cite{paper11,paper15} between the
neutrinos and the dark energy sector is the derivative interaction
between the neutrinos and the Quintessence $ \sim
{\partial_{\mu}}Q {\bar \nu_L \gamma^{\mu} \nu_L}$. During the
evolution of a homogeneous Quintessence scalar field, $\dot Q$
does not vanish which gives rise to a CPT violation in the
neutrino sector. In general, however at the present epoch $\dot Q$
is very small this type of cosmological CPT violation is predicted
to be much below the current experimental limits. But in the early
universe with high temperature it has been shown in
Ref.\cite{paper11,paper15} that this CPT violation is large enough
for the generation of the baryon number asymmetry required {\it
{via leptogenesis}}. This new mechanism for
baryogenesis/leptogenesis
 provides a
unified picture for dark energy and baryon matter of our
   Universe as I described in section III.

Another type of the interactions among neutrinos and dark energy
is the Quintessence scalar coupled to the neutrino mass term. In
the standard model of particle physics, the neutrino masses can be
described by a dimension-5 operator
\begin{equation}
  L_{\not L}=\frac{2}{f}l_{L}l_{L}\phi\phi+h.c,
\end{equation}
  where $f$ is a scale of new physics beyond the Standard Model
  which generates the $B-L$ violations, $l_{L},  \phi$ are the
  left-handed lepton and Higgs doublets respectively. When the
  Higgs field gets a vacuum expectation value $<\phi> \sim v$, the
  left-handed neutrino receives a majorana mass
  $m_{\nu} \sim \frac{v^{2}}{f}$. In Ref.\cite{paper31} we considered
an interaction between the neutrinos and the Quintessence
\begin{equation}
 \beta \frac { Q }{M_{pl}} \frac{2 }{f} l_{L}l_{L}
\phi \phi+ h.c  ,
\end{equation}
 where $\beta $ is the coefficient which characterizes the strength of the
Quintessence interacting with the neutrinos. In this scenario the neutrino
masses vary during the evolution of the universe
and we have shown that
the neutrino mass limits imposed by the baryogenesis
are modified.

The dim-5 operator above is not renormalizable, which in principle
can be generated by integrating out the heavy particles. For
example, in the model of the minimal see-saw mechanism for the
neutrino masses, we have
\begin{equation}
L=h_{ij}\bar{l}_{Li}N_{Rj}\phi+\frac{1}{2}M_{ij}\bar{N}^{c}_{Ri}N_{Rj}+h.c.
\end{equation}
 where $ M_{ij}$ is the mass matrix of
the right-handed neutrinos and the Dirac mass of neutrino is given
by $m_{D}\equiv h_{ij} <\phi> $. Integrating out the heavy
right-handed neutrinos will generate dim-5 operator, however
as pointed out in Ref.\cite{paper31}
to have the light neutrino masses varied there are various
possibilities, such as by coupling the Quintessence field to
either the Dirac masses or the majorana masses of the right-handed
neutrinos or both. In Ref.\cite{paper33} we have specifically proposed a
model
of mass
varying right-handed neutrinoes. In this model the
right-handed neutrino masses $M_i$ are assumed to be a
function of the Quintessence scalar $M_i(Q)=\overline{M}_i e^{\beta
\frac{Q}{M_{pl}}}$. Integrating out the right-handed
neutrino will generate a dimension-5 operator, but for this
case the light neutrino masses will vary in the following way
\begin{equation}
 e^{-\beta
\frac{Q}{M_{pl}}} \frac{2 }{f} l_{L}l_{L}
\phi \phi+ h.c  ,
\end{equation}
With mass varying right-handed neutrinos given above we have in
\cite{paper33} studied in detail its implication in thermal
leptogenesis. Our results show that it is possible to lower the
reheating temperature in this scenario in comparison with the case
that the neutrino masses are unchanged. This helps solve the
gravitino problem. Furthermore, a degenerate neutrino mass patten
with $m_i$ larger than the upper limit given in the minimal
thermal leptogenesis scenario is permitted.

Before concluding this section I would like to point out that the scenario
of mass varying neutrino could be tested with the Gamma Ray Burst (GRB)
by measuring the time delay\cite{paper37}
$t_d$ which for instance charactrizes the time difference
between a massive neutrino and a photon emitted from a given
source,
    \begin{equation}
     t_d\approx \int^{t_0}_t
     a(t^{\prime})dt^{\prime}\frac{1}{2}\frac{m^2}{p^2},
    \end{equation}
     where $p$ is the neutrino energy measured at the detector,
 $m$ is the
neutrino mass, and $a(t)$ is the
    expansion factor of the universe which by normalization we
set $a(t_0)=1$ at present time
    $t_0$.

From Eqs. (18) and (19) we have for the light neutrinos mass:
    \begin{equation}
     m=m_0\frac{1}{1+\beta\frac{Q_0}{M_{pl}}}(1+\beta\frac{Q}{M_{pl}})
    \end{equation}
    where $m_0$ is the
    neutrino-mass at present time, $Q_0$ is the value of Quintessence at
present
    time, and $M_{pl}$ is the Plank scale.
And for the time delay
\begin{equation}
t_d=\frac{1}{2}(\frac{m_0}{p})^2\frac{1}{(1+\beta\frac{Q_0}{M_{pl}})^2}
    \int^{t_0}_t(1+\beta\frac{Q}{M_{pl}})^2a(t^{\prime})dt^{\prime}.
 \end{equation}

\section{Summary}

Understanding the dark energy is one of the biggest challenges to the
particle
physics this century.
Studying the interaction between the dark energy and ordinary matter will
open a possibility of detecting the dark energy.
 In this paper I have shown some of the interesting
aspects of the interacting dark energy which makes connection
between the dark energy and baryo(lepto)genesis, dark energy and
dark matter, dark energy and neutrinos. Associated with the $4\%$
baryon matter there are gauge symmetry $SU_c(3) \times SU_L(2)
\times U_Y(1)$, Yukawa couplings, spontaneous symmetry breaking
.... So it might be very possible that the physics associated with
the $96 \%$ dark sector of the universe is also quite rich.

\section{Acknowledgements}

I am grateful to my collaborators and colleagues for discussions. This
work is in part supported by the national natural science foundation of
China.

\bibliographystyle{plain}

\end{document}